\documentclass[preprint,12pt]{elsarticle}

\usepackage{amssymb}
\usepackage{amsmath}

\begin{document}

\begin{frontmatter}



\title{Gravitational Effects of Idealized Electromagnetic Fields in Spherical Painleve--Gullstrand Coordinates} 


\author[addr1,addr2]{G. Abell\'an\corref{}} 
\ead{gabriel.abellan@ciens.ucv.ve, gabriel@astrumdrive.com}

\author[addr1,addr2]{N. Bol\'ivar}
\ead{nelson.e.bolivar@ucv.ve, nelson@astrumdrive.com}

\author[addr1]{I. Vasilev}
\ead{ivaylo@astrumdrive.com}



\affiliation[addr1]{
            organization={Departamento de Física, Facultad de Ciencias, Universidad Central de Venezuela},
            addressline={av. Los Ilustres}, 
            city={Caracas},
            postcode={1041-A}, 
            state={DF},
            country={Venezuela}
            }

\affiliation[addr2]{
            organization={Astrum Drive Technologies},
            addressline={Pkwy Unit 120 B, Frisco}, 
            city={Dallas},
            postcode={75034}, 
            state={Texas},
            country={USA}
            }

\begin{abstract}
We construct and analyze a class of static, spherically symmetric spacetimes in general relativity sourced exclusively by classical electrostatic configurations. Using a spherically symmetric Painlevé–Gullstrand–like metric with unit lapse and a radial shift function, we develop piecewise-defined solutions where the interior geometry is flat and the exterior is supported by various electromagnetic matter distributions. These include point-charge–like fields, Yukawa-screened electric fields, dielectric layers, and Hulth\'en-type field. The Einstein equations naturally impose a relation between the energy density and radial pressure, while the tangential pressure is derived from the metric. We systematically evaluate the classical energy conditions in each model and study the appearance of singular behavior using Israel junction conditions. This framework offers an analytically tractable setting to explore the gravitational effects of physically simple, well-understood sources without resorting to exotic matter.
\end{abstract}




\end{frontmatter}



\section{Introduction} \label{sec:intro}
Static electromagnetic field configurations in spherical geometries represent 
fundamental systems in classical electrodynamics, appearing in contexts ranging 
from charged conducting spheres to plasma confinement geometries. When such configurations carry sufficient energy density, their gravitational effects become non-negligible, requiring treatment within general relativity 
\cite{Papaetrou:1947ib,Israel:1967za,Das:1978bj,Hauser:1978vt,Copson_1978,McCrea:1982yk,Debever:1983pi,Heusler:1996ex,Kowalczynski:1998bm,Dadhich:2001sz,Sarbach:2003dz,Tod:2009em,Bonnor:2005dx,Posada:2013eqa,Manko:2013kaa,Heidmann:2023thn,Barnes:2024gko}. The resulting systems provide concrete examples of how classical electromagnetic fields serve as sources for spacetime curvature without requiring exotic matter. These configurations are constructed in a piecewise manner, featuring an interior vacuum region (described by a flat metric) and an exterior domain where the energy–momentum tensor is inspired by elementary electrostatic systems such as charged shells, dielectric layers, or screened electric fields. 
As has been discussed in \cite{Bolivar2025170147}, treating the energy density in a piecewise continuous way allows one to construct spherical configurations where the inner region is flat and the outer region is curved.

By using a Painlevé–Gullstrand–like metric with unit lapse and radial shift function $\beta(r)$, we preserve a high degree of symmetry while encoding all gravitational dynamics in the shift. The Einstein equations for this system naturally induce the equation of state $p_r=-\rho$, and the remaining components of the energy–momentum tensor depend on the specific form of the energy density $\rho(r)$. For purely electromagnetic sources, this framework yields anisotropic fluids where the tangential pressure satisfies $p_\perp=\rho$, allowing a fully analytic treatment of the system.

We focus on electrostatic matter content since it leads to well--understood, non-pathological energy–-momentum tensors that can be interpreted in terms of classical field configurations. This approach allows for explicit evaluation of the energy conditions—weak, strong, null, and dominant—under various charge distributions. 
Using these piecewise--continuous defined electrostatic sources allows us to construct regular spacetimes. This is highly guaranteed by the choice of the same type of line element to describe both the interior and exterior of the system. This method highlights how the continuity of the metric function $\beta(r)$ ensures regularity of all matter sector.

The primary goal of this paper is to construct and analyze a series of piecewise-defined gravitational solutions supported by electromagnetic sources, paying special attention to the regularity of the spacetime and the satisfaction of classical energy conditions. We examine several examples, including: a point-charge–like configuration, a Yukawa-screened field, a finite dielectric layer, and a Hulth\'en-type field. In each case, we derive the shift function, compute the full energy–momentum tensor, and assess the conditions under which the resulting spacetime is regular and physically admissible.

\section{Painleve Gullstrand Metric} \label{sec:PGmetric}
In our analysis we consider a spherically symmetric warp bubble. 
Here, we understood  a ‘warp bubble’ as a localized region of spacetime with a flat interior region and the exterior has to be asymptotically flat.
These types of configurations are studied in detail in certain astrophysical contexts \cite{Castagnino1983a,Hoye:1983dq,Nunez:1994sw,Krisch:2008bg,Cardoso:2016wcr} and also in warp drive geometries \cite{Santos-Pereira:2021mrr,Santos-Pereira:2021rsr,Santos-Pereira:2021sdm,Abellan:2023ab,Abellan:2023oym,Abellan2024a}.
This configuration will be described by a metric that closely resembles the Painlevé–Gullstrand form. The line element is
	\begin{equation}
		\label{eq:line-element}
		ds^2 = -\,dt^2 + \Bigl(dr - \beta\,dt\Bigr)^2 + r^2\,d\theta^2 + r^2\sin^2\theta\,d\varphi^2\,,
	\end{equation}
where the lapse function is set to unity and all nontrivial features of the geometry are encoded in the radial shift function \(\beta=\beta(r)\). We require that \(|\beta| < 1\) to avoid horizons. 
We adopt this coordinate system because: (i) the unit lapse concentrates all gravitational dynamics into $\beta(r)$, (ii) it provides natural treatment of piecewise geometries, and (iii) it avoids coordinate singularities 
present in other systems for these configurations.

To obtain the matter content that supports this geometry, we work with Einstein's equations
	\begin{equation}
		G_{\mu\nu} = 8\pi T_{\mu\nu}\,.
	\end{equation}
The most general energy--momentum tensor consistent with the line element \eqref{eq:line-element} corresponds to an anisotropic fluid.

\subsection{Einstein equations in a local flat frame}
In order to analyse the system of equations in a more direct way we use the local flat frame defined by the following tetrad (Cartan 1-forms)
\begin{eqnarray}\label{eq:tetrad}
    \omega^{0} = dt\,, \quad 
    \omega^1 = -\beta dt+dr\,, \quad 
    \omega^2 = rd\theta\,, \quad 
    \omega^3 = r\sin{\theta}d\varphi\,. 
\end{eqnarray}
The dual basis for this tetrad is (vector fields)
\begin{eqnarray}
    e_{0} = \partial_t + \beta\partial_r\,, \quad 
    e_1 = \partial_r\,, \quad  
    e_2 = \frac{1}{r}\partial_\theta\,, \quad 
    e_3 = \frac{1}{r\sin{\theta}}\partial_\varphi\,, \label{tetrad4}
\end{eqnarray}
with the orthonormal relation $\omega^a(e_b)=\delta^{a}_{\;\,b}$. 
In this local frame, the nonzero components of the Einstein tensor are found to be
	\begin{eqnarray}
		& & G_{00} = -G_{11} = \frac{\beta}{r^2}\Bigl(2r\,\beta' + \beta\Bigr) \,, \\[2mm]
		& & G_{22} = G_{33} = -\beta\,\beta'' - (\beta')^2 - \frac{2\beta}{r}\,\beta' = 8\pi p_\perp\,.
	\end{eqnarray}
Note that $G_{22}=G_{33}$ is fulfilled which is necessary due to the spherical symmetry. This system admits in general the description of anisotropic matter, therefore the energy-momentum tensor is given by
%
	\begin{equation}
		T_{ab} =
		\begin{pmatrix}
			\rho & 0 & 0 & 0 \\
			0 & p_r & 0 & 0 \\
			0 & 0 & p_\perp & 0 \\
			0 & 0 & 0 & p_\perp
		\end{pmatrix}\,.
	\end{equation}
From the components of the Einstein tensor $G_{ab}$, and the momentum energy tensor $T_{ab}$, we write the independent field equations
	\begin{eqnarray}
		\label{Gtt-eqn}
		& &  \frac{\beta}{r^2}\Bigl(2r\,\beta' + \beta\Bigr) = 8\pi \rho\,, \\[2mm]
		\label{Grr-eqn}
		& &  -\frac{\beta}{r^2}\Bigl(2r\,\beta' + \beta\Bigr) = 8\pi p_r\,, \\[2mm]
		\label{G22-eqn}
		 & &  -\beta\,\beta'' - (\beta')^2 - \frac{2\beta}{r}\,\beta' = 8\pi p_\perp\,.
	\end{eqnarray}
Note that from eqs.~\eqref{Gtt-eqn} and \eqref{Grr-eqn} it immediately follows that
	\begin{equation} \label{nat-eos}
		p_r(r) = -\rho(r)\,.
	\end{equation}
That is, the system envisages a naturally emerging equation of state. 

At this point, by integrating eq.~\eqref{Gtt-eqn} we obtain the following formal solution
\begin{equation}
	\label{beta-int}
	\beta^2(r) = \frac{r_0\beta^2(r_0)}{r} + \frac{8\pi}{r}\,\int_{r_0}^r \tilde{r}^2 \rho(\tilde{r})\,d\tilde{r}\,.
\end{equation}
Similarly, a comparison of \eqref{Gtt-eqn} and \eqref{G22-eqn} shows that the tangential pressure can be expressed by
\begin{equation} \label{pt-def}
	p_\perp(r)=-\frac{1}{2r}\frac{d}{dr}\Bigl[r^2\,\rho(r)\Bigr]\,.
\end{equation}
Thus, once a profile for the energy density \(\rho(r)\) is prescribed, the radial shift function \(\beta(r)\) follows from \eqref{beta-int}, and the tangential pressure \(p_\perp(r)\) is obtained via \eqref{pt-def}. These relations form the basis for our subsequent construction of warp bubble models.

\subsection{Matter Content} \label{sec:matter}
Looking at the system of equations \eqref{Gtt-eqn}--\eqref{G22-eqn} together with the equation of state \eqref{nat-eos}, we notice that we have a system of two equations with three unknowns. For this reason we must give an additional condition. The equation of state \eqref{nat-eos} which naturally appears in this system, is very restrictive. It appears frequently as an alternative to the cosmological constant in dark fluid models in cosmology. In these cases we can propose the isotropic ansatz $p_\perp=p_r=-\rho$. Substituting this ansatz into equation \eqref{pt-def} we find that
\begin{equation} \label{ansatz1-sol}
	-\rho(r)=-\frac{1}{2r}\frac{d}{dr}\Bigl[r^2\,\rho(r)\Bigr]
    \hspace{.4cm} \longrightarrow \hspace{.4cm} \rho(r) = \rho_0  \,.
\end{equation}
We note that the isotropic ansatz produces a constant energy density $\rho=\rho_0$. This can be interpreted as a simple model with constant matter density. 
This type of system can describe expansion and contraction as in some cosmological models by including the effects of dark energy within the properties of the fluid (see \cite{Huterer:2000mj,Chavanis:2015eka,Gaur:2022hap} and references therein).

However, there is a class of systems which also satisfies the equation of state $p_r = -\rho$. If the matter sector is electromagnetic, we have that equation of state and in addition we have the condition $p_\perp = \rho$. Substituting into the equation for $p_\perp$ and solving we get
\begin{equation} \label{ansatz2-sol}
	\rho(r)=-\frac{1}{2r}\frac{d}{dr}\Bigl[r^2\,\rho(r)\Bigr]
    \hspace{.4cm} \longrightarrow \hspace{.4cm} \rho(r) = \frac{Q^2}{8\pi r^4}  \,.
\end{equation}
We have chosen the constant of integration intentionally so that it is clear that this is essentially the Coulombian field. 
It is important to note that in the flat local frame defined by the tetrad \eqref{eq:tetrad}, Maxwell's equations in flat spacetime are valid, ensuring that the standard electrostatic solutions remain physically relevant sources for the Einstein equations. 
This electromagnetic ansatz is anisotropic and produces an energy density depending on $r^{-4}$ which is typical of energy densities coming from electrostatic (Coulombian) systems. 
The anisotropic pressure components, which we will discuss later, reflect the Maxwell stress tensor, where electromagnetic fields naturally exhibit directional stress with radial and tangential components differing from the isotropic perfect fluid case.
We will now consider some models driven by this electrostatic energy density. 

\section{Junction Conditions} \label{sec:coupling}
In the following sections we will be interested in describing how spacetime changes in terms of a piece-wise defined matter distribution. To determine whether thin shells form at the boundaries between regions, we employ Israel's junction condition formalism \cite{Israel1966a,Lake:1979zz,Taub:1980zr,Barrabes:1991ng}. This approach systematically examines the continuity of the induced metric and calculates the surface stress-energy tensor from discontinuities in the extrinsic curvature across matching surfaces. The Israel formalism provides the definitive criterion for identifying the presence or absence of singular matter distributions at interfaces between different spacetime regions.

\subsection{Extrinsic Curvature Calculation}
To investigate the presence of thin shells at the matching surface $r = r_0$, we employ the Israel formalism within the Cartan framework. The Israel junction conditions require the continuity of the induced metric $h_{ab}$ and determine the surface stress-energy tensor $S_{ab}$ from discontinuities in the extrinsic curvature $K_{ab}$.
In the Cartan formalism, the extrinsic curvature of the hypersurface $r = r_0$ is given by:
\begin{equation}
K_{ab} = \omega^1_{\;\;b}(e_a)\;,
\end{equation}
where $\omega^1_b$ are the spin connection 1-forms connecting the normal direction (index 1) to the tangential directions, and $e_a$ are the basis vectors tangent to the hypersurface. Using the Cartan spin connections with normal direction
\begin{align}
\omega^1_{\;\;0} = \beta'(r) w^1 \;, \quad
\omega^1_{\;\;2} = -\frac{1}{r} w^2 \;, \quad
\omega^1_{\;\;3} = -\frac{1}{r} w^3\;,
\end{align}
the non-zero components of the extrinsic curvature are
\begin{align}
K_{00} &= \omega^1_{\;\;0}(e_0) = \beta'(r) w^1(e_0) = 0\;, \label{eq:K00}\\
K_{22} &= \omega^1_{\;\;2}(e_2) = -\frac{1}{r} w^2(e_2) = -\frac{1}{r_0} \;,\label{eq:K22}\\
K_{33} &= \omega^1_{\;\;3}(e_3) = -\frac{1}{r} w^3(e_3) = -\frac{1}{r_0}\;. \label{eq:K33}
\end{align}
Note that the vanishing of $K_{00}$ in \eqref{eq:K00} follows not from the shift function $\beta(r)$ but from the geometric relation $w^1(e_0) = 0$ on the hypersurface, while the angular components in \eqref{eq:K22} and \eqref{eq:K33} depend only on the coordinate $r$ and are automatically continuous across the interface.

\subsection{Israel Junction Conditions}
To investigate the presence of thin shells at the matching surface $r = r_0$, we employ the Israel formalism within the Cartan framework. The Israel junction conditions require the continuity of the induced metric $h_{ab}$ and determine the surface stress-energy tensor $S_{ab}$ from discontinuities in the extrinsic curvature $K_{ab}$.

The induced metric on the hypersurface $r = r_0$ is obtained by restricting the spacetime metric \eqref{eq:line-element} to the surface 
\begin{equation}
h_{ab}^{\pm} = -(1-\beta_{\pm}(r_0))dt^2 + r_0^2\,d\theta^2 + r_0^2\sin^2\theta\,d\varphi^2\,,
\end{equation}
in the $(t, \theta, \phi)$ coordinates on the hypersurface. Since this expression depends only on the shift $\beta_{\pm}(r_0)$, the induced metric is continuous across the interface if
\begin{equation}
\beta_{+}(r_0) = \beta_{-}(r_0)\;.
\end{equation}
This expression is a direct consequence of the coordinate structure and ensures that the first Israel junction condition is satisfied.

The Israel formalism relates the jump in extrinsic curvature to the surface stress-energy tensor
\begin{equation}
S_{ab} = -\frac{1}{8\pi}\left([K_{ab}] - h_{ab}[K]\right)
\end{equation}
where $[K_{ab}] = K_{ab}^{+} - K_{ab}^{-}$ denotes the jump across $r = r_0$, $[K] = [K^{c}_{\;\;c}]$ is the trace of the jump, and $h_{ab}$ is the induced metric on the hypersurface.

Since all components $K_{ab}$ are either identically zero or depend only on the continuous coordinate $r$, we find:
\begin{equation}
[K_{ab}] = 0 \quad \text{for all components}
\end{equation}
This result is independent of the specific choices of $\beta_{-}(r)$ and $\beta_{+}(r)$, as long as the tetrad maintains the Painlev\'e--Gullstrand form. The vanishing of all jumps $[K_{ab}] = 0$ immediately implies
\begin{equation}
S_{ab} = 0\;.
\end{equation}
Therefore, no thin shell is required at the matching surface $r = r_0$. This remarkable result shows that the Painlev\'e--Gullstrand line element naturally accommodates the matching of interior and exterior geometries without introducing singular matter distributions at the interface.

\section{Energy Conditions}\label{sec:EC}
Energy conditions are relations between macroscopic variables that serve to assess the viability of a physical model. Certainly, many important models in physics violate some of the conditions, but it is important to consider them as guidelines for the physicality of any model in general relativity.

Here we list the weak (WEC), strong (SEC), dominant (DEC) and null (NEC) energy conditions for a general observer as already discussed in \cite{Abellan2024a}.
\subsection{Summary of general energy conditions}
For a generic timelike observer $v^{\alpha}$, a null observer $k^{\alpha}$ and a future timelike vector $F^{\alpha} = T^{\alpha}_{\;\;\beta} v^{\beta}$ in local flat frame, we find that
\begin{eqnarray}
  &   & \mbox{(WEC)} \hspace{.1cm}
    T_{\alpha \beta} v^{\alpha} v^{\beta} \geq 0\;:  
	\hspace{.7cm} 
    \rho \geq 0\,, \hspace{.4cm}
	\rho + p_i \geq 0 \,, \label{EC01}\\
  &  &\mbox{(SEC)} \hspace{.1cm}
    \left(T_{\alpha \beta}-\frac{1}{2}T\eta_{\alpha \beta}\right)v^{\alpha} v^{\beta} \geq 0\;:  
	\hspace{.5cm}  
	\rho + \sum p_i \geq 0 \,, \hspace{.2cm}
	\rho + p_i \geq 0  \,, \;\;\;\; \label{EC02} \\
   & &\mbox{(DEC)} \hspace{.1cm}
    F^{\alpha} F_{\alpha} \leq 0\;:  
	\hspace{.7cm}  
    \rho^2  \geq 0\,, \hspace{.4cm}
	\rho ^2 - p_i^2 \geq 0 \,, \label{EC03}\\
    & &\mbox{(NEC)} \hspace{.1cm}
    T_{\alpha \beta} k^{\alpha} k^{\beta} \geq 0\;:  
	\hspace{.7cm}  
	\rho + p_i \geq 0 \,, \label{EC04}
\end{eqnarray}
with $p_i$ denoting each of the principal stresses. Since Einstein's equations force equation $p_r = -\rho$ to be fulfilled, this reduces the conditions to be verified. 
Therefore, the final expressions that must be evaluated to study all the energy conditions are the following
\begin{eqnarray}\label{SimpleEC}
	\rho  \geq 0\,\,, \hspace{.4cm}
	p_\perp \geq 0 \,\,  ,\hspace{.6cm}
	\rho + p_\perp \geq 0  \,\,, \hspace{.6cm}
	\rho^2 - p_\perp^2 > 0 \,\, . 
\end{eqnarray}
This greatly simplifies the study of the energy conditions.

\section{Examples} \label{sec:examples}
In this section we consider several spherically symmetric matter configurations. First we will discuss some configurations caused by an electrostatic charge density. Then we will review some simple models with matter energy density.

\subsection{Charged shell}
Consider the following piece--wise energy density
\begin{equation} \label{eq:density-ex1}
    \rho(r)\;=\; \left\{
	\begin{array}{ll}
		0, & 0 \le r<R_1,\\
		\displaystyle \frac{Q^2}{8\pi r^4}, & r\ge R_1.
	\end{array} \right.
\end{equation}
Such an energy density could be caused by a homogeneous charge distribution in a spherical shell of radius $R_1$. In this case, the electric field cancels out inside the surface ($r<R_1$) and corresponds to a point charge on the outside ($r\geq R_1$). Recall that for this type of electrostatic systems, $\rho = E^2/8\pi$ is satisfied. To find the shift function we use the expression $\eqref{beta-int}$ for each region.

\subsubsection{Region I: $0\leq r < R_1$.}
Using that $r_0 = 0$ and $\rho(r)=0$, one has directly $\beta_{-}(r)=0$.

\subsubsection{Region II: $r\geq R_1$.}
\begin{figure*}[htbp]
\centering
\begin{tabular}{cccc}
\hspace{-.5cm}
\includegraphics[width=0.24\textwidth]{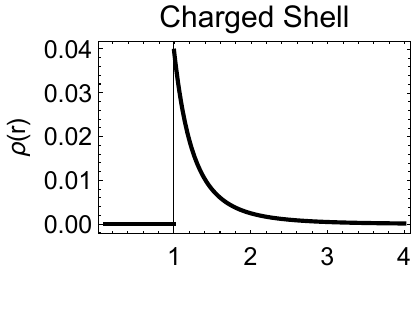} & \hspace{-.5cm}
\includegraphics[width=0.24\textwidth]{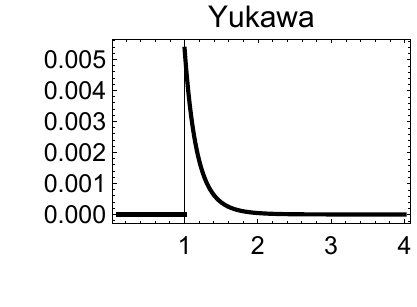} &  \hspace{-.5cm}
\includegraphics[width=0.24\textwidth]{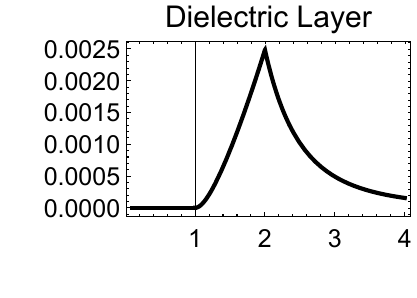} & \hspace{-.5cm}
\includegraphics[width=0.24\textwidth]{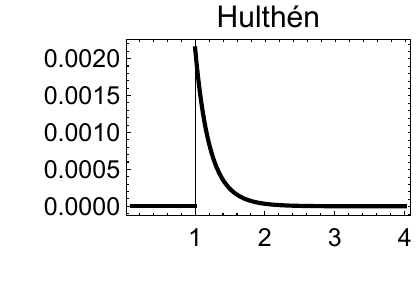} \\
\hspace{-.5cm}
\includegraphics[width=0.24\textwidth]{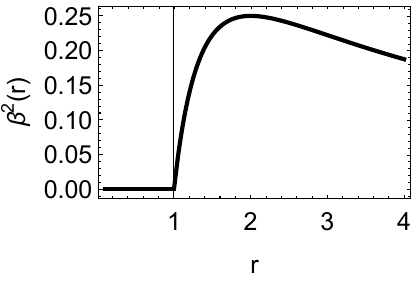} & \hspace{-.5cm}
\includegraphics[width=0.24\textwidth]{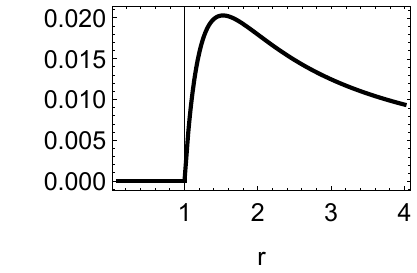} & \hspace{-.5cm}
\includegraphics[width=0.24\textwidth]{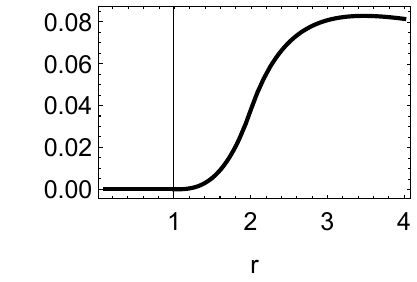} & \hspace{-.5cm}
\includegraphics[width=0.24\textwidth]{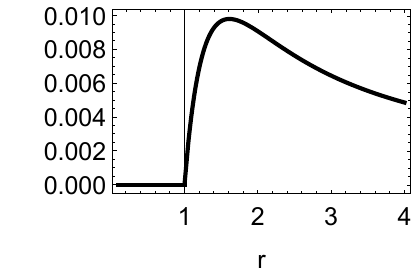} \\
\end{tabular}
\caption{(Top) Electromagnetic energy density profiles $\rho(r)$, (bottom) shift functions $\beta^2(r)$. The piecewise structure is evident: flat interior regions ($r < R_1$) and electromagnetically sourced mid and exterior regions. Parameters: $Q^2 = 8\pi$, $R_1 = 1$, $R_2 = 2$ (dielectric), $k = 1$ (screening), $B^2 = 8\pi$ (Hulth\'en).}
\label{fig:electromagnetic_profiles}
\end{figure*}

Now we use that $r_0=R_1$. By imposing that $\beta$ is continuous in $R_1$, in this case the solution is written as
\begin{eqnarray}
	\label{beta-ex1}
	\beta^2_+(r) &=& \frac{R_1\beta^2(R_1)}{r} + \frac{8\pi}{r}\,\int_{R_1}^r \tilde{r}^2 \rho(\tilde{r})\,d\tilde{r} \nonumber \\
    &=& \frac{Q^2}{R_1^2} \Bigg[ \bigg( \frac{R_1}{r} \bigg) - \bigg( \frac{R_1}{r} \bigg)^2 \Bigg]  \,.
\end{eqnarray}
Clearly, we have that the shift function $\beta$ is continuous in $R_1$ (see figure \ref{fig:electromagnetic_profiles}). 
Using this shift function, we can check that the tangential pressure is given by the expression $p_\perp(r) = \rho(r)$.

\subsubsection{Energy conditions}
To evaluate the energy conditions we must use the expressions \eqref{SimpleEC} and note that given that \eqref{eq:density-ex1} with relations $p_r(r)=-\rho(r)$ and $p_\perp(r)=\rho(r)$ it follows that all the energy conditions are satisfied, provided that $\rho(r)\geq0$ is satisfied. 
It is worth commenting that although the phenomenological picture of this problem starts from considering a thin shell of charge uniformly distributed on a spherical surface of radius $R_1$, the energy--momentum tensor associated to this distribution is perfectly regular and does not produce any thin energy/matter shells, which is remarkable.

\subsection{Yukawa type field}
Let us now consider a Yukawa type field. This type of field can arise as an effective electric field in theories where charge shielding occurs. The electric field we are going to study is written as
\begin{equation}\label{yukawa-field}
    E(r) = \frac{Q}{r^2}e^{-kr} \;,
\end{equation}
where the parameter $k>0$ measures the effective shielding length. Such configurations occur naturally in plasmas where a cloud of electrons shields the positive charge on the nucleus. Given this expression for the electric field, we write the energy density
\begin{equation} \label{eq:density-ex2}
    \rho(r)\;=\; \left\{
	\begin{array}{ll}
		0, & 0 \le r<R_1,\\
		\displaystyle \frac{Q^2}{8\pi r^4}e^{-2kr}, & r\ge R_1.
	\end{array} \right.
\end{equation}
Let us now analyse the behaviour of the shift function for each region.

\subsubsection{Region I: $0\leq r < R_1$.}
Using that $r_0 = 0$ and $\rho(r)=0$ in \eqref{beta-int}, one has $\beta_{-}(r)=0$. So in the inner region, a flat spacetime is obtained, and therefore the material sector corresponds to the vacuum.

\subsubsection{Region II: $r\geq R_1$.}
Now we use that $r_0=R_1$ and by imposing that $\beta$ is continuous in $R_1$ we have $\beta_+(R_1)=0$, in this case the solution is written as
\begin{eqnarray}
	\label{beta-ex2}
	\beta^2_+(r) &=& \frac{R_1\beta^2(R_1)}{r} + \frac{8\pi}{r}\,\int_{R_1}^r \tilde{r}^2 \rho(\tilde{r})\,d\tilde{r} \nonumber \\
    &=& \frac{Q^2}{r} \Big[ F_2(R) - F_2(r) \Big]\,. \nonumber
\end{eqnarray}
Here we have defined the function $F_p(w)$ as
\begin{equation}\label{exp-int-func}
    F_p(w)=\frac{e^{-p k w}}{w}+pk\,\text{Ei}(-pk w)\,, \quad\quad
    \text{Ei}(x) = -\int_{-x}^\infty dt\; t^{-1} e^{-t}\;. 
\end{equation}
with $\text{Ei}(x)$ the exponential integral function. 
It is worth noting a few points. First, taking the limit $k\to 0$ in $F_p(w)$ recovers the charged shell solution already studied. Secondly, that the shift function $\beta$ is continuous in $R_1$.

Once we have the metric function $\beta(r)$, we can calculate the material sector. Given the natural equation of state, we see that the radial pressure is $p_r(r)=-\rho(r)$ with $\rho(r)$ taken as the expression \eqref{eq:density-ex2}. Finally we calculate the tangential pressure and find that
\begin{equation} \label{eq:pt-ex2}
    p_\perp(r)\;=\; \left\{
	\begin{array}{ll}
		0, & 0 \le r<R_1,\\
		\displaystyle \frac{Q^2}{8\pi r^4}e^{-2kr}(1 + kr), & r\ge R_1.
	\end{array} \right.
\end{equation}
It is not surprising that the tangential pressure does not correspond exactly to the energy density, since the electric field is not exactly that corresponding to the static Coulombian field. This suggests that the shielding produces shear stresses in the material distribution. Finally, we check that in the limit $k\to 0$ the solution for the previously studied charged shell is recovered.

\subsubsection{Energy conditions}
For region I all energy conditions yield trivial relations. Let's look at what occurs in region II. Here we have that $\rho>0$ and given that $k>0$, we also have that $p_\perp >0$ is satisfied. Therefore we see that these three inequalities are satisfied $\rho \geq 0$, $p_\perp \geq 0$, and $\rho + p_\perp \geq 0$. 

Finally let us analyse the expression $\rho^2 - p^2_\perp \geq 0$. Notice that we can write $p_\perp = \rho (1+kr)$. This means that we have
\begin{eqnarray}
    \rho^2 - p^2_\perp = \rho^2[1 - (1+kr)^2] \geq 0
    \hspace{.3cm} \longrightarrow \hspace{.3cm} 1 - (1+kr)^2 \geq 0 \;.
\end{eqnarray}
We see that this inequality cannot be fulfilled. This implies that the dominant energy condition is not satisfied.

We have examined the model for the Yukawa type field, and found that the weak, null and strong energy conditions are fulfilled, leaving the dominant energy condition unsatisfied.

\subsection{Dielectric layer}
Our next example starts from the problem of considering a uniform charge density distributed in a spherical layer of radii $R_1$ and $R_2$. The electric field that solves this electrostatic problem is
\begin{equation} \label{eq:electric-ex3}
    E(r)\;=\; \left\{
	\begin{array}{ll}
		0\;, & 0 \le r<R_1,\\
		\displaystyle \frac{Q}{r^2} \bigg(
        \frac{r^3-R_1^3}{R^3_2-R^3_1}\bigg)  , & R_1 \leq r < R_2\;,\\
        \displaystyle \frac{Q}{r^2}\;, & r \geq R_2\;.
	\end{array} \right.
\end{equation}
Using this expression for the electric field, we write the energy density
\begin{equation} \label{eq:density-ex3}
    \rho(r)\;=\; \left\{
	\begin{array}{ll}
		0\;, & 0 \le r<R_1,\\
		\displaystyle \frac{Q^2}{8\pi r^4} \bigg(
        \frac{r^3-R_1^3}{R^3_2-R^3_1}\bigg)^{\!2}  , & R_1 \leq r < R_2\;,\\
        \displaystyle \frac{Q^2}{8\pi r^4}\;, & r \geq R_2\;.
	\end{array} \right.
\end{equation}
Now we analyse the behaviour of the shift function $\beta$ for each region.

\subsubsection{Region I: $0\leq r < R_1$.}
In this region one has that the metric function $\beta_I(r)$ vanishes. This corresponds to the flat space and is therefore a trivial vacuum solution.

\subsubsection{Region II: $R_1 \leq r < R_2$.}
By using the expression \eqref{beta-int} with $r_0=R_1$ and requiring it to be continuous at this point, we have that the shift function is written as
\begin{eqnarray}`
	\label{beta-ex3-b}
	\beta^2_{II}(r) &=& \frac{R_1\beta^2_{II}(R_1)}{r} + \frac{8\pi}{r}\,\int_{R_1}^r \tilde{r}^2 \rho(\tilde{r})\,d\tilde{r} \nonumber \\
    &=& \frac{Q^2}{r^2} \Bigg[  
    \frac{r^6 - 5R_1^3 r^3+9R_1^5r-5R_1^6}{5(R^3_2-R_1^3)^2}\Bigg] \,.
\end{eqnarray}
When we evaluate this expression in $r=R_1$, it cancels out $\beta^2_{II}(R_1)=0$, as must be the case for continuity. On the other hand, we evaluate the limit when $r\to R_2$, so that we can impose continuity when evaluating junction with region III
\begin{equation} \label{beta-R2-ex3}
    \beta_{III}^2(R_2) =  \lim_{r \to R_2} \beta^2_{II}(r) =
    \frac{Q^2}{R_2^2} \Bigg[  
    \frac{R_2^6 - 5R_1^3 R_2^3+9R_1^5R_2-5R_1^6}{5(R^3_2-R_1^3)^2}\Bigg] \,.
\end{equation}
Using the expression \eqref{beta-ex3-b} for the metric function $\beta^2_{II}(r)$ we can calculate the matter sector in this region. We see that the radial pressure is $p_{II}^{(r)}=-\rho_{II}$ as given by \eqref{eq:density-ex3}. 
Finally, the tangential pressure $p^{(\perp)}_{II}$ is given by
\begin{eqnarray} \label{pt-ex3-b}
	p_{II}^{(\perp)}(r)
    = \frac{Q^2}{8\pi (R^3_2-R_1^3)^2 r^4} (R_1^3-r^3) (2r^3+R_1^3) \,,
\end{eqnarray}
Note that in the region where the dielectric is present, condition $p_{II}^{(\perp)}(r) = \rho_{II}(r)$ is not satisfied. Analogous to the Yukawa-type case, the presence of the dielectric appears to indicate the existence of shear stresses.

\subsubsection{Region III: $r \geq R_2$.}
We Consider equation \eqref{beta-int} with $r_0=R_2$. By requiring $\beta(r)$ to be continuous at this point, the shift function is written as
\begin{eqnarray}
	\label{beta-ex3-c}
	\beta^2_{III}(r) &=& \frac{R_2\beta^2_{III}(R_2)}{r} + \frac{8\pi}{r}\,\int_{R_2}^r \tilde{r}^2 \rho(\tilde{r})\,d\tilde{r} \nonumber \\
    &=& \frac{R_2\beta^2_{III}(R_2)}{r} +   
    \frac{Q^2}{r^2}\Big(\frac{r}{R_2}-1\Big) \,,
\end{eqnarray}
In this case, we note that the value when evaluating $\beta^2_{III}(R_2)$ is given by the limit calculated in \eqref{beta-R2-ex3}. This ensures that the metric function is continuous in $R_2$.

By using equation \eqref{beta-ex3-c} we calculate the matter sector for the exterior region. The radial pressure is $p_{III}^{(r)}=-\rho_{III}$ as shown by \eqref{eq:density-ex3}. 
Furthermore, the tangential pressure in this region is $p^{(\perp)}_{III}(r)=\rho_{III}(r)$ 
\begin{eqnarray} \label{pt-ex3-c}
	p_{III}^{(\perp)}(r)
    = \frac{Q^2}{8\pi r^4}\;,
\end{eqnarray}
this corresponds to the charged shell solution.

\subsubsection{Energy conditions}
Since region I has the flat vacuum solution, the energy conditions are trivially satisfied. Similarly in region III, for $r>R_2$, all the conditions \eqref{SimpleEC} are also satisfied since the solution is the one corresponding to the Coulombian field.

Let's take a closer look at what occurs in region II. From expression \eqref{eq:density-ex3}, we see that $\rho_{II}\geq 0$ is satisfied. Let us now consider the behaviour of tangential pressure $p_{II}^{(\perp)}$. In this region one has
\begin{eqnarray} \label{pt-EC-ex3}
	p_{II}^{(\perp)}(r)
    =  \frac{Q^2}{8\pi (R^3_2-R_1^3)^2 r^4} (R_1^3-r^3) (2r^3+R_1^3)
    \geq 0 
    \hspace{.3cm} \longrightarrow \hspace{.3cm}
    R_1^3-r^3 \geq 0
    \,.
\end{eqnarray}
It can be clearly seen that this condition is not satisfied, since by construction we have $r>R_1$.

Now we examine condition $\rho_{II} + p_{II}^{(\perp)}\geq 0$. As can be seen from \eqref{eq:density-ex3} and \eqref{pt-ex3-b}, in region II one has
\begin{eqnarray} \label{rhopt-EC-ex3}
	\rho_{II} + p_{II}^{(\perp)}(r)
    \geq 0 
    \hspace{.3cm} \longrightarrow \hspace{.3cm}
    2R_1^6 - R_1^3 r^3 - r^6 \geq 0
    \,.
\end{eqnarray}
Let us analyse for $f(r)=2R_1^6 - R_1^3 r^3 - r^6$, formally we can see that this function satisfies $f(r)>0$ for value $r^3 \in (-2R_1^3,R_1^3)$. Hence, we have $\rho_{II} +p_{II}^{(\perp)} < 0$ and the inequality is not satisfied.

The last condition to be examined is $(\rho_{II})^2 - (p_{II}^{(\perp)})^2 \geq 0$. This condition can be written as
\begin{eqnarray} \label{rho2pt2-EC-ex3}
	(\rho_{II})^2 - (p_{II}^{(\perp)})^2 \geq 0 
    \hspace{.3cm} \longrightarrow \hspace{.3cm}
    5 r^{12} - 8 R_1^3 r^9 + 3 R_1^6 r^6 - 2 R_1^9 r^3 + 2 R_1^{12} \geq 0
    \,.
\end{eqnarray}
Defining the function $f(r)=5 r^{12} - 8 R_1^3 r^9 + 3 R_1^6 r^6 - 2 R_1^9 r^3 + 2 R_1^{12}$, we can check that the only real root is given by $r^3=R_1^3$. Therefore this function fulfils $f(r)\geq 0$ and therefore the condition $(\rho_{II})^2 - (p_{II}^{(\perp)})^2 \geq 0$ is satisfied in region II.

We have examined the energy conditions for a dielectric layer-type matter distribution. In doing so we see that in regions I and III the energy conditions are fully met. However, for region II the system is very restrictive and satisfies only conditions $\rho+ p_r \geq 0$ and $\rho^2 - p_r^2 \geq 0$ (which are trivials) in addition to $\rho \geq 0$ and $\rho^2 - p_\perp^2 \geq 0$. Thus only the dominant energy condition \eqref{EC03} is satisfied. The rest of the conditions are violated in the region where the hypothetical dielectric medium is located.

\subsection{Hulthén Field}\label{sec:hulthen}
In this last example, the matter sector is characterized by an energy density taken as the square of a field:
\begin{equation}\label{eq:density-ex4}
	\rho(r)=
	\begin{cases}
		0\,, & r < R_1\,,\\[1mm]
		\dfrac{B^2}{8\pi r^4}e^{-2 k r}\bigl(1-e^{-k r}\bigr)^2 \,, & r\ge R_1\,,
	\end{cases}
\end{equation}
where \(B>0\) is an amplitude and \(k>0\) is a decay parameter. 
Next we calculate the metric function for each region.

\subsubsection{Region I: $0\leq r < R_1$.}
In the inner region, a flat spacetime is obtained $\beta_{-}(r)=0$, and consecuently the material sector corresponds to the vacuum.

\subsubsection{Region II: $r\geq R_1$.}
From equation \eqref{beta-int} we use that $r_0=R_1$ and by imposing that $\beta$ is continuous in $R_1$ we have $\beta_+(R_1)=0$, in this case the solution is written as
\begin{eqnarray}
	\label{beta-ex4}
	\beta^2_+(r) &=& \frac{R_1\beta^2(R_1)}{r} + \frac{8\pi}{r}\,\int_{R_1}^r \tilde{r}^2 \rho(\tilde{r})\,d\tilde{r} \nonumber \\
    &=& \frac{B^2}{r} \Big[ \big( F_4(R_1)-F_4(r) \big) - 2\big( F_3(R_1)-F_3(r) \big) + \big( F_2(R_1)-F_2(r) \big)\Big]. \;\;
\end{eqnarray}
Here we use the $F_p(w)$ function defined in \eqref{exp-int-func}.

We now proceed to calculate the tangential pressure. Using the expression \eqref{G22-eqn} (or \eqref{pt-def}), we obtain 
\begin{equation} \label{eq:pt-ex4}
    p_\perp(r)\;=\; \left\{
	\begin{array}{ll}
		0\;, & r<R_1,\\
		\displaystyle \frac{B^2}{8\pi r^4} e^{-4kr}
    (e^{kr}-1)[e^{kr}(1+kr) - (1+2kr)]\;, & r\ge R_1.
	\end{array} \right.
\end{equation}
We can see that this case resembles the Yukawa-type case already analysed. In this case we have again that $p_{+}^{(\perp)}(r) \neq \rho_{+}(r)$: the tangential pressure cannot satisfy the simple condition of a charged shell.

\subsubsection{Energy conditions}
Region I is a flat vacuum solution, so the energy conditions are trivially satisfied there.

We focus at what occurs in region II. From expression \eqref{eq:density-ex4}, we see that $\rho_{II}\geq 0$ is satisfied for all $r\geq R_1$. Let us now examine the tangential pressure $p_{II}^{(\perp)}$. In this region one should has
\begin{eqnarray} \label{pt-EC-ex4-a}
	p_{II}^{(\perp)}(r)
    = \frac{B^2}{8\pi r^4} e^{-4kr}
    (e^{kr}-1)[e^{kr}(1+kr) - (1+2kr)]
    \geq 0 
    \,.
\end{eqnarray}
Since in this region we have $B^2 e^{-4kr}(e^{kr}-1)/8\pi r^4 > 0$, then to satisfy the inequality, it suffices that the following is satisfied
\begin{eqnarray} \label{pt-EC-ex4-b}
    e^{kr}(1+kr) - (1+2kr) \geq 0
    \,.
\end{eqnarray}
Recall that $kr=x>0$ in region II. Defining a function $f(x)=e^{x}(1+x) - (1+2x)$ and noting that $f\to 0$ when $x\to 0$, we can see that this function is a strictly increasing function on $x \geq 0$.
This shows that this function satisfies $f(r)>0$ and consequently $p_{II}^{(\perp)} \geq 0$ is satisfied.

Next, we examine condition $\rho_{II} + p_{II}^{(\perp)}\geq 0$. From \eqref{eq:density-ex4} and \eqref{eq:pt-ex4}, in region II one has
\begin{eqnarray} \label{rhopt-EC-ex4}
	\rho_{II} + p_{II}^{(\perp)}(r)
    \geq 0 
    \hspace{.3cm} \longrightarrow \hspace{.3cm}
    e^{kr}(2+kr) - 2(1+kr) \geq 0
    \,.
\end{eqnarray}
Similar to the previous case, we define a function $f(x)=e^{x}(2+x) - 2(1+x)$ and noting that $f\to 0$ when $x\to 0$, we can see that this function is a strictly increasing function $f'(x)>0$ on $x\geq 0$.
As a result, this function satisfies $f(r)>0$ and consequently $\rho_{II} + p_{II}^{(\perp)} \geq 0$ is satisfied.

The last condition is $(\rho_{II})^2 - (p_{II}^{(\perp)})^2 \geq 0$. This condition can be written as
\begin{eqnarray} \label{rho2pt2-EC-ex4}
	(\rho_{II})^2 - (p_{II}^{(\perp)})^2 \geq 0 
    \hspace{.3cm} \longrightarrow \hspace{.3cm}
    (e^{kr}-2)[2(1+kr) - e^{kr}(2+kr)]  \geq 0
    \,.
\end{eqnarray}
Defining the function $f(x)=(e^{x}-2)[2(1+x) - e^{x}(2+x)] = A(x) B(x)$, it is clear that $A(x)$ and $B(x)$ must have the same sign in order to satisfy the last inequality. Now, $B(x)$ has already been analysed for condition \eqref{rhopt-EC-ex4}, only this time it gives us a strictly decreasing function since it is multiplied by -1, hence $B(x) < 0$. It only remains to examine $A(x)=e^{x}-2$. It can easily be shown that $A(x)<0$ when you have $x<\ln{2}$. But this does not cover the entire region II, for this reason the condition $(\rho_{II})^2 - (p_{II}^{(\perp)})^2 \geq 0$ is not met.

Having carefully examined the energy conditions, we see that in region I all conditions are totally met while in region II the conditions $\rho \geq 0$, $p_\perp \geq 0$ and $\rho+ p_\perp \geq 0$ are fulfilled. Thus the weak \eqref{EC01}, null \eqref{EC04} and strong \eqref{EC02} energy conditions are satisfied. The dominant energy condition are violated in the region II where the Hulth\'en medium is located.


\section{Discussion}
The electromagnetic configurations studied here represent idealized versions of systems that appear throughout classical physics. Spherical charge distributions model conducting spheres and plasma boundaries; screened fields describe systems with effective screening mechanisms such as Debye shielding in plasmas; dielectric layers correspond to multilayer spherical capacitors or optical systems; and modified field profiles capture deviations from pure Coulomb behavior perhaps due to finite-size effects or material responses.

The four electromagnetic configurations presented in Section \ref{sec:examples} show a coherent and technically rich picture of how classical electric fields interact with gravity within spherically symmetric, piecewise-defined spacetimes. The analytical control granted by the Painlevé–Gullstrand–like metric and the separation between a flat interior and an exterior matter region permits a direct mapping between the profile of the electric field and the geometric structure of spacetime. 

One aspect to note is that for these models, there are no thin shells. This is a consequence of the chosen metric and the boundary conditions imposed on the function $\beta(r)$. The analysis of the Israel conditions clearly showed that the material sector is regular for all models presented in this work.

A notable feature of the system is the intrinsic anisotropic character of the energy–momentum tensor. The relation $p_r=-\rho$, which emerges directly from the Einstein equations within the metric ansatz, strongly constrains the possible forms of matter. For electromagnetic sources examined, the tangential pressure satisfies $p_\perp = \rho$ only in the case of pure Coulomb fields, while screened or spatially modulated fields break this balance and introduce shear stresses.

The energy condition analysis reveals a consistent pattern:
\begin{itemize}
    \item The point charge model satisfies all classical energy conditions throughout the spacetime. This is expected, as the energy density and pressures mirror those of a pure Coulomb field in electrovacuum.
    \item The Yukawa and Hulth\'en models, while regular, violate the dominant energy condition (DEC) due to the modified anisotropy produced by the radial modulation of the electric field.
    \item The dielectric layer model satisfies only the dominant condition, but violates the weak, strong, and null conditions within the dielectric region. This underscores how dielectric screening, though physically motivated, can give rise to configurations that are not fully compatible with all classical energy constraints.
\end{itemize}
These observations indicate that violation of energy conditions does not necessarily entail singular behavior, nor does it imply the need for exotic sources. Rather, in this context, it signals a high degree of anisotropy in the matter content, as seen in screened field configurations. This distinguishes these systems from standard perfect-fluid models and highlights the rich structure of electromagnetic energy–momentum tensors.

Each model analyzed provides a controlled setting to isolate specific physical or geometrical effects. The charge shell illustrates the canonical behavior of charged vacuum spacetimes and serves as a baseline for benchmark. By comparing these models, we observe that the onset of energy condition violations correlates with increasing spatial localization and anisotropy of the field. This suggests that, while global regularity is preserved, the detailed structure of the energy–momentum tensor becomes increasingly constrained as more physically realistic (i.e., non-Coulombic) behaviors are introduced.

\section{Final Remarks}
The models presented in this work offer a concrete and analytically tractable realization of how classical electromagnetic fields shape the geometry of spacetime in general relativity, without the need for speculative or exotic matter sources. By constructing spherically symmetric, piecewise-defined spacetimes with flat interiors and electromagnetic exteriors, we have been able to probe the fine structure of energy–momentum distributions and their gravitational manifestations with clarity. It is important to note that by using the Cartan formalism and moving to the local flat frame, we can use the results found in known solutions of Maxwell's equations and thus study their effects on the spacetime using Einstein's equations.

One of the main conceptual outcomes of this study is the realization that even highly idealized systems—such as a charged shell, a dielectric layer, or a screened electric field—can produce nontrivial and physically meaningful spacetime geometries. Remarkable, these geometries are regular. 
The role of the radial shift function $\beta(r)$, as the carrier of all nontrivial geometric information, provides a powerful analytical handle for constructing and understanding these configurations.
From a physical perspective, the restriction to a flat interior is not merely a simplification, but rather a modeling choice that ensures the gravitational features observed arise solely from the presence of electromagnetic fields in the exterior region. This is particularly relevant when the models are interpreted as approximations to systems with charged boundaries, cavities, or hollow structures, where the core is effectively empty. 

The piecewise techniques employed here are relevant for broader applications in theoretical and mathematical physics. They could be useful in the modeling of gravitational interfaces, domain walls, multiphase media, where different regions of spacetime are governed by distinct physical laws or field configurations. The methodology may also find a place in semiclassical gravity, where quantum fields with localized energy densities induce backreaction on the background geometry. We believe that these aspects are worth exploring in future work.

\bibliographystyle{spphys}


\bibliography{biblio.bib}
	











\end{document}